\documentclass[aps, physrev, twocolumn, preprintnumbers,10pt]{revtex4-2}
\usepackage{graphicx}
\usepackage{dcolumn}
\usepackage{bm}
\usepackage{amsmath}
\usepackage{amssymb}
\usepackage{multirow}
\usepackage{subcaption}
\usepackage{caption}
\usepackage[colorlinks=true, allcolors=blue]{hyperref}

\renewcommand{\arraystretch}{1.5}
\begin{document}
	\title{SU(3) flavor symmetry analysis of hyperon non-leptonic two body decays }

  \author{Xin Wu$^{1,2}$}
    \author{Qi Chen$^{1,2}$}
      \author{Ye Xing$^{3}$}
      \email{xingye_guang@cumt.edu.cn}
	\author{Zhi-Peng Xing$^{1,2}$}
	\email{zpxing@nnu.edu.cn}
	\author{Ruilin Zhu$^{1,2}$}
    \email{rlzhu@njnu.edu.cn}

    \affiliation{$^1$Department of Physics and Institute of Theoretical Physics, Nanjing Normal University, Nanjing, Jiangsu 210023, China}
     \affiliation{$^2$Nanjing Key Laboratory of Particle Physics and Astrophysics}
     \affiliation{$^3$School of Physics,China University of Mining and Technology, Xuzhou 221000, China}

		\begin{abstract}
    This paper present a systematic study of hyperon non-leptonic two-body decays induced by light quark transitions, particularly the $s \rightarrow u\bar{u}d$ process, within the framework of SU(3) flavor symmetry. The effective weak Hamiltonian is decomposed into irreducible SU(3) representations, including the 27-plet and octet components, and applied to analyze decays of octet and decuplet baryons and charmed baryons. Both the irreducible representation amplitude (IRA) approach and the topological diagrammatic analysis (TDA) are employed to construct decay amplitudes and constrain the parameter space. SU(3) symmetry-breaking effects arising from the strange quark mass are incorporated systematically. A global fit to current experimental data allows us to extract form factors and predict branching ratios and asymmetry parameters for several decay channels, including $\Lambda^0 \rightarrow p\pi^-$, $\Sigma^+ \rightarrow p\pi^0$, and $\Omega^- \rightarrow \Xi^0\pi^-$. Our results demonstrate the predictive power of SU(3) flavor symmetry while highlighting significant symmetry-breaking effects, especially in amplitudes related to the 27-plet. Notably, the $\Sigma^+ \rightarrow p\pi^0$ decay channel exhibits a deviation exceeding $1\sigma$ from experimental measurements, suggesting the possible presence of new decay mechanisms or contributions beyond the Standard Model. This work provides a systematic framework for future tests of the Standard Model and the search for new physics in hyperon decays.
	\end{abstract}

	\maketitle

	\section{Introduction}
	Weak decays of hyperons have long constituted a crucial issue for testing the Standard Model (SM) and exploring new physics (NP). From the experimental side, hyperons, as intermediate states in heavy hadron cascade decay processes, are abundantly produced due to their low production threshold at many experimental facilities such as BESIII and LHCb, etc.~\cite{BESIII:2017hyw,LHCb:2017rdd,LHCb:2020jpq,BESIII:2022qax,BESIII:2022rgl,BESIII:2023fhs,BESIII:2023utd}. Recently, BESIII has new measurements for the absolute branch fraction of $\Omega^{-}\to \Xi^0\pi^{-} $, $\Omega^{-}\to \Xi^-\pi^{0} $, $\Omega^{-}\to \Lambda^0 K^{-} $ decay, and the sensitivity is now in the range of $10^{-5}\sim10^{-8}$~\cite{BESIII:2023ldd}.
	From the theoretical side, hyperon weak decays involve a large CKM matrix element $V_{us}V_{ud}^*$. The precise testing of this CKM matrix element are helpful for testing the unitarity of the CKM matrix. Besides, due to its low threshold,it contains rich non-perturbative QCD effects and different decay behavior which are reflected in their complex angular distributions~\cite{He:2022jjc}. In recent years, there has been much theoretical work focusing on hyperon decays such as the structure of hyperons~\cite{Gockeler:2002uh}, decay mechanism\cite{Xing:2023jnr}, angular distribution\cite{Fu:2023ose} and so on.
	
	As for the theoretical study of the hyperon decays, the strict factorization cannot work well due to the low threshold and the smaller transform energy in the s quark decay modes. Therefore the perturbative calculation  is currently not feasible. Besides the  perturbative study, model calculation\cite{Dubovik:2008zz,Hu:2018luj,Zenczykowski:2020hmg,Wang:2022tcm,Xing:2023jnr}, effective theory\cite{Zenczykowski:1999vq,Shi:2022dhw,Shi:2023kbu,Han:2023hgy,Shi:2025xkp}, symmetry analysis\cite{Wang:2019alu,Xu:2020jfr,Niu:2020aoz}, Lattice QCD~\cite{Bali:2024oxg,LatticeParton:2024vck} and so on \cite{Chang:2000hu,Li:2016tlt,He:2023cqg} are still worked well in hyperon decays. However due to the large model independence, the uncertainty in model calculation is difficult to estimate\cite{Xing:2023jnr}. The effective theory such as Chiral perturbation theory($\chi$PT) has been very successful in studying low-energy strong interaction physics. However, it may need deeper study on the puzzle of $\alpha(\Sigma^+\to p \gamma)$\cite{Shi:2023kbu}. As a symmetry analysis method although the SU(3) analysis does not involve the detailed dynamic understanding, it seems can solve this problem\cite{Zenczykowski:2005cs}. Therefore the analyze the hyperon decays in SU(3) symmetry is useful.
	
	In symmetry analysis, although it is not possible to calculate the absolute values of the decay amplitudes, it can obtain relations between different decay amplitudes.  With a smaller number of amplitudes, when combined with experimental data, the amplitudes can be constrained and predictions can be made to further test the approach  without a detail understanding of the dynamics. Recently, there are many symmetry analysis work on the hyperon decay processes\cite{Wang:2019alu,Xu:2020jfr,Niu:2020aoz}. As an important part of hyperon decays, the hyperon non-leptonic two body decays have accumulated a large amount of experimental data and rich Phenomenological observables which have attracted the interest of theorists. In these processes, the 
	symmetry analysis such as isospin symmetry which reflect to up and down quark symmetry have previous studied~\cite{Wang:2019alu}. Though the isospin symmetry is powerful in hyperon non-leptonic two body weak decay processes, the SU(3) symmetry reflect to u, d, s symmetry which will bring less parameter and  will provide more information such as the CPV. We notice that the preliminary SU(3) topological diagram analysis is given~\cite{Xu:2020jfr}. Unfortunately, the SU(3) symmetry analysis basic on the strict irreducibility represents decomposition is still absent in these processes. Therefore the further SU(3) symmetry analysis on hyperon non-leptonic two body decay processes is necessary and urgent.
	
	The rest of this paper is organized as follows. In Sec.II the theoretical framework of hyperon non-leptonic two body weak decay under SU(3) symmetry are given. The decomposition of Hamiltonian which is the $3 \otimes 3 \otimes \overline{3} \otimes \overline{3}$ SU(3) group representation are derived in the first time. 
	In Sec.III the octet light baryon two body decays are studied under SU(3) symmetry. Then the decuplet light baryon two body decays are also analyzed in Sec.IV. Induced by  the same Hamiltonian, the charmed baryon and octet light meson decay can also be studied which are given in Sec.V and Sec. VI respectively. A conclusion is given in the last section.
	
	\section{Theoretical framework}
	
Under SU(3) symmetry, baryons composed of light quarks can be classified into an octet and a decuplet, while light mesons form an octet. To study the hyperon two body weak decays, the effective Hamiltonian is given as\cite{Buchalla:1995vs}
	\begin{eqnarray}
		{\cal H}_{eff}=\frac{G_F}{\sqrt{2}}V_{ud}V_{us}^*\sum^{10}_{i=1}\bigg[z_i(\mu)-\frac{V_{td}V_{ts}^*}{V_{ud}V_{us}^*}y_i(\mu)\bigg]Q_i(\mu),
	\end{eqnarray}
	where $V_{uq}$ is the CKM matrix element and the $z_i$ and $y_i$ are the Wilson coefficients. For the current-current operator $Q_{1/2}$, $y_{1/2}$ is equal to zero. For the penguin operator $Q_{3-10}$, we have $z_{3-10}=0$. For studying the primary contribution of the Hamiltonian, the tree level operator $Q_{1}$ and $Q_2$ are only considered in our work. These specific expressions of four-quark operator $Q_{1/2}$ are
	\begin{eqnarray}
		&Q_1=[\bar d_\alpha u_\beta]_{V-A}[\bar u_\beta s_\alpha]_{V-A},\notag\\
        &Q_2=[\bar d_\alpha u_\alpha]_{V-A}[\bar u_\beta s_\beta]_{V-A}.\label{tree}
	\end{eqnarray}
	By extracting the flavor information, the tree level Hamiltonian can be redefined as 
	\begin{eqnarray}
		{\cal H}_{eff}&=&\frac{G_F}{\sqrt{2}}V_{ud}V_{us}^*\notag\\
		&\times&\sum_{\lambda=1,2} C_\lambda\sum_{i,j,k,l}(H_\lambda)^{ij}_{kl}[\bar q^i_\alpha q^k_{\alpha}]_{V-A}[\bar q^j_\beta q^l_{\beta}]_{V-A},\label{ham}\notag\\
	\end{eqnarray}
	where $q^{1}$, $q^2$ and $q^3$ correspond to the u, d and s quark, respectively. The Wilson coefficient $C_{\lambda}$ is $z_\lambda(\mu)-\frac{V_{td}V_{ts}^*}{V_{ud}V_{us}^*}y_\lambda(\mu)$. The matrix $H_\lambda$ only contain two nonzero elements as $(H_2)^{21}_{13}=1$ and $(H_1)^{12}_{13}=1$ with $\lambda=1,2$.
	
	In SU(3) irreducibility representation amplitude (IRA) method, the matrix $H_\lambda$ can be seen as the $3 \otimes 3 \otimes \overline{3} \otimes \overline{3}$ SU(3) group representation and it can be composed as $3 \otimes 3 \otimes \overline{3} \otimes \overline{3} = 27 \oplus 10\oplus \overline{10} \oplus 8 \oplus 8 \oplus 8 \oplus 8 \oplus 1 \oplus 1$. In preliminary analysis, it is known that the representation $27$ and  $(10$, $\overline {10})$ should be the symmetric traceless representation as $(H_{27})^{\{ij\}}_{\{kl\}}$, $(H_{10})^{\{ijm\}}$ and $(H_{\overline{10}})_{\{ijm\}}$. Both the  trace in these representation are absorbed into $8$ and $1$. For constructing these representation matrices, the Hamiltonian matrix $H^{ij}_{kl}$ can be symmetrized and anti-symmetrized to four terms as
	\begin{eqnarray}
		H^{ij}_{kl}&=&H^{\{ij\}}_{\{kl\}}+H^{\{ij\}}_{[kl]}+H^{[ij]}_{\{kl\}}+H^{[ij]}_{[kl]},\notag\\
		H^{\{ij\}}_{\{kl\}}&=&\frac{1}{4}(H^{ij}_{kl}+H^{ji}_{kl}+H^{ij}_{lk}+H^{ji}_{lk}),\notag\\
		H^{\{ij\}}_{[kl]}&=&\frac{1}{4}(H^{ij}_{kl}+H^{ji}_{kl}-H^{ij}_{lk}-H^{ji}_{lk}),\notag\\
		H^{[ij]}_{\{kl\}}&=&\frac{1}{4}(H^{ij}_{kl}-H^{ji}_{kl}+H^{ij}_{lk}-H^{ji}_{lk}),\notag\\
		H^{[ij]}_{[kl]}&=&\frac{1}{4}(H^{ij}_{kl}-H^{ji}_{kl}-H^{ij}_{lk}+H^{ji}_{lk}).\label{HD}
	\end{eqnarray} 
	One can directly find that $H^{\{ij\}}_{\{kl\}}$ contain the  $27$ representation. For constructing the $10$ and $\overline{10}$ representation, one can also define the matrix as
	\begin{eqnarray}
		H^{\{ij\}m}=\epsilon^{mkl}H^{\{ij\}}_{[kl]},H_{\{kl\}n}=\epsilon_{ijn}H^{[ij]}_{\{kl\}}.
	\end{eqnarray}  
	One can see that only index i j are symmetry in $H^{\{ij\}m}$ and k l are symmetry in $H_{\{kl\}n}$.
	After the symmetrized and anti-symmetrized processing, the completely symmetry matrix can be constructed as
	\begin{eqnarray}
		H^{\{ij\}m}&=&H^{\{i\{j\}m\}}+H^{\{i[j\}m]}\notag\\
		&=&\frac{3}{2}H^{\{ijm\}}-\frac{1}{2}H^{\{mj\}i}+\frac{1}{2}\epsilon^{jmn}\mathring{H}^i_n,\notag\\
		H_{\{kl\}n}&=&H_{\{k\{l\}n\}}+H_{\{k[l\}n]}\notag\\
		&=&\frac{3}{2}H_{\{kln\}}-\frac{1}{2}H_{\{kl\}n}++\frac{1}{2}\epsilon_{lnm}\dot{H}^m_k,
	\end{eqnarray}  
	with
	\begin{eqnarray}
		\mathring{H}^i_n=\epsilon_{njm}H^{\{i[j\}m]},\;\dot{H}^m_k=\epsilon^{mln}H_{\{k[l\}n]}.
	\end{eqnarray}      
	Then we can solve that 
	\begin{eqnarray}
		H^{\{ijm\}}&=&H^{\{ij\}m}-\frac{1}{9}\epsilon^{ijk}\mathring{H}^m_n-\frac{4}{9}\epsilon^{jmn}\mathring{H}^i_n+\frac{2}{9}\epsilon^{min}\mathring{H}^j_n,\notag\\
		H_{\{kln\}}&=&H_{\{kl\}n}-\frac{1}{9}\epsilon_{klm}\dot{H}^m_n-\frac{4}{9}\epsilon_{lnm}\dot{H}^m_k+\frac{2}{9}\epsilon_{nkm}\dot{H}^m_l.\notag\\
	\end{eqnarray}    
	For the last term in Eq.\eqref{HD}, we can extract the octet $8$ and singlet $1$ as
	\begin{eqnarray}
		H^{[ij]}_{[kl]}&=&\frac{1}{2}\epsilon^{ijm}\epsilon_{kln} \ddot{H}^n_m+\frac{1}{6}(\delta^i_k\delta^j_l-\delta^i_l\delta^j_k)\overline{H},\notag\\
		\ddot{H}^n_m&=&\frac{1}{2}\epsilon_{ijn}\epsilon^{klm}H^{[ij]}_{[kl]}-\frac{1}{3}\delta^m_nH^{[ij]}_{[ij]},\;\overline{H}=H^{[ij]}_{[ij]},
	\end{eqnarray} 
	where the $\ddot{H}^n_m$ reflect to octet is traceless and its trace can be singlet. Following the same method, the $H^{\{ij\}}_{\{kl\}}$ can be composed into $27$ ,$8$ and $1$ as
	\begin{eqnarray}
		H^{\{ij\}}_{\{kl\}}&=&\tilde H^{\{ij\}}_{\{kl\}}+\frac{1}{5}(\delta^i_k \hat H^j_l+\delta^j_k\hat H^i_l+\delta^i_l\hat H^j_k+\delta^j_l\hat H^i_k)\notag\\
		&&+\frac{1}{12}(\delta^i_k\delta^j_l+\delta^i_l\delta^j_k)H,
	\end{eqnarray} 
	where $\tilde H^{\{ij\}}_{\{kl\}}$ is traceless. The irreducibility representation can be given by the redefinition  as 
	\begin{eqnarray}
		&&\tilde H^{\{ij\}}_{\{kl\}}\to (H_{27})^{\{ij\}}_{\{kl\}}, \;\; H^{\{ijm\}}\to(H_{10})^{ijm},\notag\\
		&&H_{\{kln\}}\to(H_{\overline{10}})_{kln},\;\;\hat H^m_n\to(H^1_8)^m_n,\;\;\mathring{H}^m_n\to (H^2_8)^m_n,\notag\\
		&&\dot{H}^m_n\to (H^3_8)^M_n,\;\;\ddot{H}^m_n\to (H^4_8)^m_n,\;\;H\to H^1_1,\notag\\
		&&\overline{H}\to H^2_1.
	\end{eqnarray}  
	Therefore, finally, the Hamiltonian matrix $H^{ij}_{kl}$ can be decomposed by irreducibility representations  as
	\begin{eqnarray}
		H^{ij}_{kl}&=&(H_{27})^{\{ij\}}_{\{kl\}}+\frac{1}{2}\epsilon_{klm}(H_{10})^{ijm}+\frac{1}{2}\epsilon^{ijn}(H_{\overline{10}})_{kln}\notag\\
		&&+\frac{1}{5}A^{ijn}_{klm}(H^1_8)^m_n-\frac{1}{6}B^{ijn}_{klm}(H^2_8)^m_n-\frac{1}{6}C^{ijn}_{klm}(H^3_8)^m_n\notag\\
		&&+\frac{1}{2}\epsilon^{ijn}\epsilon_{klm}(H^4_8)^m_n+\frac{1}{12}(\delta^i_k\delta^j_l+\delta^i_l\delta^j_k)H^1_1\notag\\
		&&-\frac{1}{6}(\delta^i_k\delta^j_l-\delta^i_l\delta^j_k)H^2_1,
	\end{eqnarray}
	with
	\begin{eqnarray}
		A^{ijn}_{klm}&=&\delta^i_k\delta^j_m\delta^n_l+
		\delta^j_k\delta^i_m\delta^n_l+
		\delta^i_l\delta^j_m\delta^n_k+
		\delta^j_l\delta^i_m\delta^n_k,\notag\\
		B^{ijn}_{klm}&=&\delta^i_k\delta^j_m\delta^n_l+
		\delta^j_k\delta^i_m\delta^n_l-
		\delta^i_l\delta^j_m\delta^n_k-
		\delta^j_l\delta^i_m\delta^n_k,\notag\\
		C^{ijn}_{klm}&=&\delta^i_k\delta^j_m\delta^n_l-
		\delta^j_k\delta^i_m\delta^n_l+
		\delta^i_l\delta^j_m\delta^n_k-
		\delta^j_l\delta^i_m\delta^n_k.
	\end{eqnarray}
	Then these irreducibility representations can be extracted as
	\begin{eqnarray}
		(H_{27})^{\{ij\}}_{\{kl\}}&=&H^{\{ij\}}_{\{kl\}}\notag-\frac{1}{5}A^{ijn}_{klm}(H^1_8)^m_n-\frac{1}{12}(\delta^i_k\delta^j_l+\delta^i_l\delta^j_k)H^1_1,
		\notag\\
		(H_{10})^{ijm}&=&H^{\{ij}_{kl}\epsilon^{m\}kl},\;\; (H_{\overline{10}})_{kln}=\epsilon_{ij\{n}H^{ij}_{kl\}},\notag\\
		(H^1_8)^m_n&=&H^{\{im\}}_{\{in\}}-\frac{1}{3}\delta^m_nH^{\{ij\}}_{\{ij\}},\notag\\
		(H^2_8)^m_n&=&H^{\{mj\}}_{nj}-H^{jm}_{\{nj\}}, \;
		(H^3_8)^m_n=H^{mj}_{\{nj\}}-H^{jm}_{\{nj\}},\notag\\
		(H^4_8)^m_n&=&\frac{1}{2}\epsilon_{ijn}\epsilon^{klm}H^{[ij]}_{[kl]}-\frac{1}{3}\delta^m_n H^{[ij]}_{[ij]},\notag\\
        H^1_1&=&H^{\{ij\}}_{\{ij\}},\;\; H^2_1=H^{[ij]}_{[ij]}.\label{decom}
	\end{eqnarray}
	After the decomposition, one can derive the specific expression of each irreducibility representations by using the input Hamiltonian matrix. Since the singlet is  trivial, we will omit them in our following analysis.
	However, we notice that the information for the Hamiltonian may provide further constrain on specific irreducibility representations. In the first step, we can redefine the Hamiltonian in Eq.\eqref{ham} as $H_+=(H_1+H_2)/2$ and $H_-=(H_1-H_2)/2$. The Hamiltonian can be expressed as 
    \begin{eqnarray}
		{\cal H}_{eff}&=&\frac{G_F}{\sqrt{2}}V_{ud}V_{us}^*\notag\\
		&\times&\sum_{\lambda=+,-} C_\lambda\sum_{i,j,k,l}(H_\lambda)^{ij}_{kl}[\bar q^i_\alpha q^k_{\alpha}]_{V-A}[\bar q^j_\beta q^l_{\beta}]_{V-A},\notag\\\label{hampm}
	\end{eqnarray}
    where $C_{\pm}=C_1\pm C_2$.
    One can see that the index in $H_+$ is symmetry and $H_-$ is anti-symmetry. This indicate that the $H_+$  contain the $H_{27}$, $H_{10}$, $H_{\overline{10}}$ and  $H^{1,2,3}_8$ irreducibility representations.  The matrix $H_-$ only contain the $H^4_8$. We also find that the Hamiltonian in Eq.\eqref{ham} shows the symmetry of upper index in $H_\lambda$ are related to the lower index. It can be shown as 
	\begin{eqnarray}
		[\bar q^i_\alpha q^k_{\alpha}]_{V-A}[\bar q^j_\beta q^l_{\beta}]_{V-A}=[\bar q^j_\beta q^l_{\beta}]_{V-A}[\bar q^i_\alpha q^k_{\alpha}]_{V-A}.
	\end{eqnarray}
	This suggest that $(H_\lambda)^{ij}_{kl}=(H_\lambda)^{ji}_{lk}$. Then we have
	\begin{eqnarray}
		H^{\{ij\}}_{[kl]}&=&\frac{1}{4}(H^{ij}_{kl}+H^{ji}_{kl}-H^{ij}_{lk}-H^{ji}_{lk})=0,\notag\\
		H^{[ij]}_{\{kl\}}&=&\frac{1}{4}(H^{ij}_{kl}-H^{ji}_{kl}+H^{ij}_{lk}-H^{ji}_{lk})=0.
	\end{eqnarray}
	Therefore the $H_{10}$, $H_{\overline{10}}$, $H^2_8$ and $H^3_8$ are equal to zero in our work.
	\section{The  octet light baryon two body decays}
    With the help of the Hamiltonian decomposition, we are able to analyze the octet baryon two body decays $T_8 \rightarrow T_8 P_8$ with IRA method. The $T_8$ and $P_8$ respectively represent the light baryon octet and the pseudoscalar meson octet.
They can be written as:
	\begin{eqnarray}\
		T_8&=&
		\begin{pmatrix}
			\frac{\Sigma^0}{\sqrt{2}}+\frac{\Lambda}{\sqrt{6}}& \Sigma^+ &p \\
			\Sigma^- & -\frac{\Sigma^0}{\sqrt{2}}+\frac{\Lambda}{\sqrt{6}}&n\\
			\Xi^-& \Xi^0&-\frac{2\Lambda}{\sqrt{6}}
		\end{pmatrix},\notag\\
		P&=&
		\begin{pmatrix}
			\frac{\pi^0+\eta_q}{\sqrt{2}}& \pi^+ &K^+ \\
			\pi^- & \frac{-\pi^0+\eta_q}{\sqrt{2}}&K^0\\
			K^-& \bar{K}^0&\eta_s
		\end{pmatrix}.
	\end{eqnarray}
	Following the analysis in Sec.II, the tree operators in Eq.\eqref{tree} can be decomposed under the SU(3) flavor  symmetry as
	$3 \otimes 3 \otimes \overline{3} \otimes \overline{3} = 27 \oplus 10\oplus \overline{10} \oplus 8 \oplus 8 \oplus 8 \oplus 8 \oplus 1 \oplus 1$ and only $27$ and $8$ have nonzero contributions.
	
	For the specific processes induced by $s\rightarrow u\bar{u}d$, the Hamiltonian matrices are $(H_{+})^{12}_{13}=(H_{+})^{21}_{13}=\frac{1}{2}$ and $(H_{-})^{12}_{13}=-(H_{-})^{21}_{13}=\frac{1}{2}$. The representation of the IRA  Hamiltonian are
	\begin{align}
		&H_{27\{13\}}^{\{12\}} = H_{27\{31\}}^{\{21\}} = H_{27\{31\}}^{\{12\}} = H_{27\{13\}}^{\{21\}} = \frac{1}{5} \sin \theta,\nonumber \\
		&H_{27\{23\}}^{\{22\}} = H_{27\{32\}}^{\{22\}} = H_{27\{33\}}^{\{23\}} = H_{27\{33\}}^{\{32\}} = -\frac{1}{10} \sin \theta,\nonumber \\
		&{H_{8}}^2_3 = \frac{1}{4} \sin \theta.\label{deha}
	\end{align}
	Since the $H^1_8$ and $H^2_8$ have the same contribution for $s\rightarrow u\bar{u}d$ transition, we only use the $H^8$ to express these two octets. 
	With the above expressions, one may derive the effective Hamiltonian for decays involving the  octet baryon as
	\begin{align}
		\mathcal{M}_{T_8 \rightarrow T_8P_8}&=a_{27}(T_{8})_i^j(H_{27})_{\left\{jk\right\}}^{\left\{il\right\}}(T_8)_l^mP_m\nonumber\\
		&+b_{27}(T_{8})_i^j(H_{27})_{\left\{jk\right\}}^{\left\{im\right\}}(T_8)_l^kP_m^l\nonumber\\
		&+c_{27}(T_{8})_i^j(H_{27})_{\left\{jk\right\}}^{\left\{lm\right\}}(T_8)_l^iP_m^k\nonumber\\
		&+d_{27}(T_{8})_i^j(H_{27})_{\left\{jk\right\}}^{\left\{lm\right\}}(T_8)_l^kP_m^i\nonumber\\
		&+e_{27}(T_{8})_i^l(H_{27})_{\left\{jk\right\}}^{\left\{im\right\}}(T_8)_l^jP_m^k\nonumber\\
		&+f_{27}(T_{8})_i^m(H_{27})_{\left\{jk\right\}}^{\left\{il\right\}}(T_8)_l^jP_m^k\nonumber\\
		&+a_{8}(T_8)_i^j(H_8)_j^k(T_8)_k^lP_l^i\nonumber\\
		&+b_{8}(T_8)_i^j(H_8)_j^l(T_8)_k^iP_l^k\nonumber\\
		&+c_{8}(T_8)_i^k(H_8)_j^l(T_8)_k^jP_l^i\nonumber\\
		&+d_{8}(T_8)_i^l(H_8)_j^i(T_8)_k^jP_l^k\nonumber\\
		&+e_{8}(T_8)_i^l(H_8)_j^k(T_8)_k^jP_l^i,
	\end{align}
where $a_{27}\sim f_{27}$ and $a_8\sim e_8$ are SU(3) irreducible amplitudes. The expression shows that the amplitude of octet baryon two body decays $T_8 \rightarrow T_8 P_8$ can be expressed by these 11 parameters.  To determine these parameters, we need to use the experimental data as given in Table.\ref{table1}. Unfortunately, the current data are insufficient to determine these parameters. Since each SU(3) irreducible amplitude can be divide  into parity conserving and parity violating term, the total number of the IRA method parameter is 22 which is larger than the number of the observables. Therefore, in this work, the number of SU(3) irreducible amplitude is not counted correctly.

\begin{figure}[h!]
		\centering
		\includegraphics[width=1\linewidth]{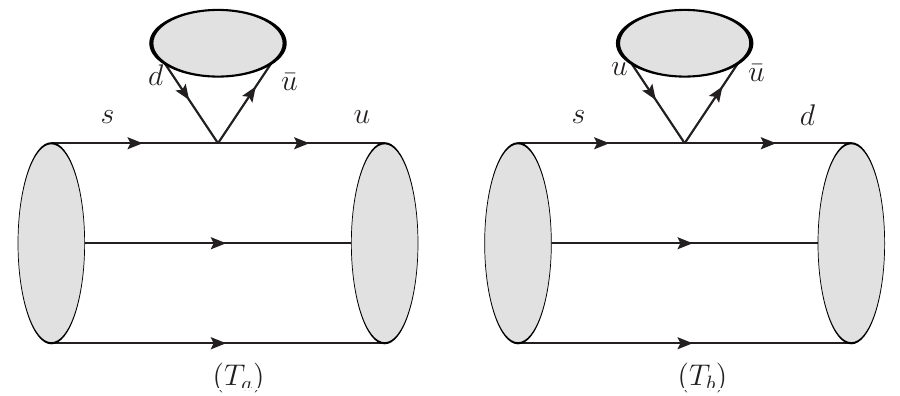}
		\caption{Color symmetric topological diagrams for $T_8 \rightarrow T_8P_8$  nonleptonic decays.}
		\label{fig:symmetry}
	\end{figure}

By including the color information in the Hamiltonian matrix, one can find that the two quark field of anti-quark field in  $H^{\{ij\}}_{\{kl\}}$ is color symmetric and in $H^{[ij]}_{[kl]}$ is anti-symmetric. Since the color must be anti-symmetric in baryon state, the amplitude in which the initial/final baryon state directly connect the two quark/anti-quark in Hamiltonian is expected to be strongly suppressed\cite{Korner:1970xq,Pati:1970fg}. 
To correctly count the number of SU(3) irreducible amplitudes, we can use the topological diagrammatic approach(TDA) method to give an intuitive physical image. We find that for the color symmetric IRA Hamiltonian, only two topological diagrams (Fig.\ref{fig:symmetry}) can contribute. Then the number of amplitudes of $H_{27}$ and $H_8^1$ can be largely reduced.
Since the equivalence between the TDA and irreducible SU(3) methods has been verified\cite{He:2018php,He:2018joe}, we convert the amplitude obtained by the TDA amplitude into IRA, and only write down the IRA amplitude which include the two topological diagram contribution in Fig.\ref{fig:symmetry}. We find that only two amplitudes $c_{27}$ and $e_{27}$ corresponding to the two topological diagrams. The explicit expressions of these amplitudes are given as follows:
\begin{align}
T_a:(T_8)^{[in]j}H_{mn}^{kl}(T_8)_{[ik]j}P_l^m&=-(T_{8})_i^l(H_{27})_{\left\{jk\right\}}^{\left\{im\right\}}(T_8)_l^jP_m^k\nonumber\\
(T_8)^{[in]j}H_{mn}^{kl}(T_8)_{[kj]i}P_l^m&=(T_{8})_i^j(H_{27})_{\left\{jk\right\}}^{\left\{lm\right\}}(T_8)_l^iP_m^k\nonumber\\
&+(T_{8})_i^l(H_{27})_{\left\{jk\right\}}^{\left\{im\right\}}(T_8)_l^jP_m^k\nonumber\\
T_b:(T_8)^{[in]j}H_{mn}^{kl}(T_8)_{[il]j}P_k^m&=-(T_{8})_i^l(H_{27})_{\left\{jk\right\}}^{\left\{im\right\}}(T_8)_l^jP_m^k\nonumber\\
(T_8)^{[in]j}H_{mn}^{kl}(T_8)_{[ij]l}P_k^m&=(T_{8})_i^j(H_{27})_{\left\{jk\right\}}^{\left\{lm\right\}}(T_8)_l^iP_m^k\nonumber\\
&+(T_{8})_i^l(H_{27})_{\left\{jk\right\}}^{\left\{im\right\}}(T_8)_l^jP_m^k.
    \end{align}
We find that the contribution of $e_{27}$ in the two amplitudes contained in the topology $T_a $($T_b$) is opposite. As a result, they cancel each other when the amplitudes are summed, leaving only the contribution of $c_{27}$. 
Based on this simplification, the IRA amplitudes can be written as: 
	\begin{align}
		\mathcal{M}_{T_8 \rightarrow T_8P_8}&=c_{27}(T_{8})_i^j(H_{27})_{\left\{jk\right\}}^{\left\{lm\right\}}(T_8)_l^iP_m^k\nonumber\\
		&+a_{8}(T_8)_i^j(H_8)_j^k(T_8)_k^lP_l^i\nonumber\\
		&+b_{8}(T_8)_i^j(H_8)_j^l(T_8)_k^iP_l^k\nonumber\\
		&+c_{8}(T_8)_i^k(H_8)_j^l(T_8)_k^jP_l^i\nonumber\\
		&+d_{8}(T_8)_i^l(H_8)_j^i(T_8)_k^jP_l^k\nonumber\\
		&+e_{8}(T_8)_i^l(H_8)_j^k(T_8)_k^jP_l^i.\label{M8}
	\end{align}
	We notice that although the amplitude of $H_8^1$ can be largely reduced, the $H_8^4$ which comes from the $H^{[ij]}_{[kl]}$ will not be constrained. Therefore the number of $H_8$ amplitudes remains 5. Form these amplitude, we can get the relation of the amplitude by expanding them according to each decay channel in Table.\ref{table1} as
    \begin{align}
        \mathcal{M}(\Sigma^{0}\to p  \pi^{-} )=-\mathcal{M}(\Sigma^{+}\to p\pi^{0}).\label{sr}
    \end{align}
    In eq.\eqref{M8}, one can see that there are 6 such amplitudes. In fact, these amplitudes can be expressed by 12 form factors. Generically, we can express them by parity conserving and parity violating as
	\begin{eqnarray}
		q_{27}&=&G_F\bar{u}(f^c_{27} - g^c_{27}\gamma_5)u,\notag\\
		q_{8}&=&G_F\bar{u}(f^q_{8} - g^q_{8}\gamma_5)u,\quad q=a,b,c,d,e.\label{ff}
	\end{eqnarray}
	For the process of octet baryon two body
	decays $T_8 \rightarrow T_8P_8$, the non-polarization decay width is easily written as
	\begin{eqnarray}
		\frac{d\Gamma}{d\cos\theta_M}=\frac{G_{F}^{2}|\vec{p}_{B_{n}}|(E_{B_n}+M_{B_{n}})}{8\pi M_{B_s}}(|F|^2+\kappa^2 |G|^2),\label{decaywidth}\notag\\
	\end{eqnarray}
	where $\hat p_{B_n}$ is final state momentum. Depending on the specific processes, the $F$ and $G$ linear functions of $f^i_{8/27}$ and $g^i_{8/27}$ are the scalar and peseudoscalar form factors, respectively.
	The parameter $\kappa$ writing in terms of masses is given by
	\begin{eqnarray}
		\kappa=|\Vec{p}_{B_n}|/(E_{B_n}+M_{B_n}),
		\kappa^2 = {(M_{B_s} - M_{B_n})^2 - M^2_M \over (M_{B_s} + M_{B_n})^2 - M^2_M}.\notag\\
	\end{eqnarray}
	Here $M_{B_s}$, $M_{B_n}$, and $M_M$ are the masses of  the initial octet baryons, final octet baryons, and mesons, respectively.

 In this step, we find that the six SU(3) parameters can not explain the experimental data well. However, these results are not unexpected. The hyperon decays usually involve large SU(3) symmetry breaking considering the mass of strange quark. Therefore, in our analysis, the symmetry breaking should be considered.

Since the mass of strange quark is much larger than the up and down quark as $m_s$ $\gg$ $m_{u,d}$, the SU(3) symmetry breakdown induced by  the different masses of the light u, d and s quarks can be introduced.  The quark mass matrix can be written as 
	\begin{eqnarray}\
		M &=&
		\begin{pmatrix}
			m_u & 0 & 0 \\
			0 & m_d & 0 \\
			0 & 0 & m_s
		\end{pmatrix}
		\sim m_u
		\begin{pmatrix}
			1 & 0 & 0 \\
			0 & 1 & 0 \\
			0 & 0 & 1
		\end{pmatrix}
		+ m_s
		\begin{pmatrix}
			0 & 0 & 0 \\
			0 & 0 & 0 \\
			0 & 0 & 1
		\end{pmatrix}\nonumber\\
		&=& m_u
		\begin{pmatrix}
			1 & 0 & 0 \\
			0 & 1 & 0 \\
			0 & 0 & 1
		\end{pmatrix}
		+ m_s \times \omega.
	\end{eqnarray}

	The first matrix represents the mass term under the strict SU(3) symmetry  and the second matrix can be seen as interaction term which represents SU(3) symmetry  breaking effect. Based on this mass matrix, the SU(3) symmetry-breaking contributions to the irreducible representation amplitudes can be constructed using the interaction matrix $\omega$ as:
    \begin{align}
		\mathcal{M}^{SB}_{T_8 \rightarrow T_8P_8}&=\mathbf{c^1_{27}}(T_8)_i^n(H_{27})_{jk}^{lm}(T_8)_l^iP_m^k\omega_n^j\notag\\
        &+\mathbf{c^2_{27}}(T_8)_i^j(H_{27})_{jk}^{lm}(T_8)_l^nP_m^k\omega_n^i\notag\\
        &+\mathbf{a_8}(T_8)_i^n(H_8)_j^k(T_8)_k^lP_l^i\omega_m^j\notag\\
        &+\mathbf{b^1_8}(T_8)_i^j(H_8)_j^l(T_8)_k^mP_l^k\omega_m^i\notag\\
        &+\mathbf{b^2_8}(T_8)_i^m(H_8)_j^l(T_8)_k^iP_l^k\omega_m^j\notag\\
        &+\mathbf{c^1_8}(T_8)_i^m(H_8)_j^l(T_8)_k^jP_l^i\omega_m^k\notag\\
        &+\mathbf{c^2_8}(T_8)_i^k(H_8)_j^l(T_8)_k^mP_l^i\omega_m^j\notag\\
        &+\mathbf{d_8}(T_8)_i^l(H_8)_j^i(T_8)_k^mP_l^k\omega_m^j\notag\\
        &+\mathbf{e_8}(T_8)_i^l(H_8)_j^k(T_8)_k^mP_l^i\omega_m^j.    
    \end{align}
    Although 9 SU (3) symmetry breaking terms were established, we found that the degree of freedom of our IRA amplitude are only 7.
    
   Since the decay information expressed in the SU(3) symmetry breaking terms $\mathbf{c^1_{27/8}}$, $\mathbf{a_8}$, $\mathbf{d_8}$, $\mathbf{e_8}$ and $\mathbf{b^1_8}$ are the same as that expressed in corresponding SU(3) symmetry terms, we can absorb these symmetry breaking contribution into the amplitude in Eq.\eqref{M8}.  Finally, we found that there would only be three SU (3) symmetry-breaking terms remaining.
	\begin{align}
		\mathcal{M}^{SB}_{T_8 \rightarrow T_8P_8}&=\mathbf{c^2_{27}}(T_8)_i^j(H_{27})_{jk}^{lm}(T_8)_l^nP_m^k\omega_n^i\notag\\
        &+\mathbf{b^2_8}(T_8)_i^j(H_8)_j^l(T_8)_k^mP_l^k\omega_m^i\notag\\
        &+\mathbf{c^2_8}(T_8)_i^m(H_8)_j^l(T_8)_k^jP_l^i\omega_m^k.
	\end{align}
 Then one can  define new parameters $A_8$, $B_8$ to absorb the symmetry breaking term $\mathbf{b^2_8}$ and $\mathbf{c^2_8}$ as $A_8=a_8-\mathbf{b^2_8}-2\mathbf{c^2_8}$, $B_8=b_8+\mathbf{b^2_8}$. After the redefinition, the degrees of freedom are 8. Then the amplitude  $\mathbf{c_{27}}$ only contains the symmetry breaking contribution and other amplitude are contributed by symmetry and its breaking together. Therefore, the value of $\mathbf{c_{27}}$ will represent the contribution of symmetry breaking effect.
 
 By the least-$\chi^2$ fit method\cite{Chen:2025drl}, we can give a global analysis on these processes. 
 
In our work, we find that the SU(3) IRA amplitude including symmetry breaking can basically explain the experimental except the $Br(\Sigma^{+}\to p  \pi^{0} )$ and $\alpha(\Sigma^{+}\to p  \pi^{0} )$ have more than 1 $\sigma$ deviation. By expanding the error of these two data to 2 $\sigma$, we can achieve a reasonable $\chi^2/d.o.f=1.54$. The fitting results and prediction results are given in Table.\ref{table1} and the fitting parameters are given in Table. \ref{parameters}. 

Our results show that the puzzle of $Br(\Sigma^{+}\to p  \pi^{0} )$ and $\alpha(\Sigma^{+}\to p  \pi^{0} )$ can not be explained by SU(3) symmetry breaking effect. It suggested that this channel may including new decay mechanism including the new physics contribution. We also look forward to future experiments that can collect more data such Super Tau-Charm Facility(STCF) to reduce the data error and determine this puzzle. We also suggest the  theorist can pay more attention on this process. 
\begin{table}[h]
	\caption{SU(3) symmetry breaking irreduciable amplitudes from fitting.}\label{parameters}
	\begin{tabular}{|c|c|c|c|c|c|c|c|c|c|c|}\hline\hline
		\multirow{2}{*}{parameters}&\multicolumn{2}{c|}{ Our work }\cr\cline{2-3}
		&scalar(f) & pseudoscalar(g)\\\hline
        $A_{8}$  & $-3.20\pm0.45$&$-1.74\pm0.46$\\\hline
		$B_{8}$  & $-3.32\pm0.45$&$-10.49\pm0.45$\\\hline
		$c_{8}$  & $-2.39\pm0.45$&$-8.84\pm0.45$\\\hline
        $d_{8}$  & $-3.04\pm0.45$&$-2.64\pm0.45$\\\hline
        $e_{8}$  & $2.42\pm0.45$&$2.48\pm0.45$\\\hline
		$c_{27}$  & $0.0672\pm0.0010$&$0.031\pm0.088$\\\hline
		$\mathbf{c^2_{27}}$  & $-0.0503\pm0.0032$&$-0.23\pm0.12$\\\hline
		$\chi^{2}$/d.o.f&\multicolumn{2}{c|}{1.54} \cr\cline{1-3}
	\end{tabular}
\end{table}

\begin{table*}
	\centering
	\renewcommand{\arraystretch}{1.5}
	\caption{The irreducible representation amplitudes (second column), the experimental data (third column) and predicted values (fourth column) of the decay branching ratio, and the experimental values (fifth column) and predicted values (sixth column) of the asymmetric parameters regarding the hyperon two-body non-leptonic decay.}\label{table1}
	\begin{tabular}{|c|c|c|c|c|c|c|c|}
		\hline
		\multirow{2}{*}{Channel} & \multirow{2}{*}{Amplitudes}&\multicolumn{2}{c|}{Br($10^{-2}$)} & \multicolumn{2}{c|}{$\alpha$} \\
		\cline{3-6}
		&& Experiment data &Our work& Experiment data &Our work \\
		\hline
		$\Sigma^{0}\to p  \pi^{-} $&$ \frac{\text{$\sin \theta$} \left(5 c_8-5 d_8\right)}{20 \sqrt{2}}$&-&$4.713(22)\times10^{-8}$&-&$-0.99917(20)$\\\hline
		$\Sigma^{0}\to n  \pi^{0} $&$ \frac{\text{$\sin \theta$} \left(5 c_8+5 d_8+10 e_8\right)}{40}$&-&$2.340(12)\times10^{-8}$&-&$0.99620(47)$\\\hline
		$\Lambda^{0}\to p  \pi^{-} $&$ \frac{\text{$\sin \theta$} \left(-10 b_8+5 c_8-8 c_{27}+5d_8-10\mathbf{b^2_8}-8\mathbf{c^2_{27}}\right)}{20 \sqrt{6}}$&$64.1(5)$&$64.1(5)$&$0.747(9)$&$0.747(9)$\\\hline
		$\Lambda^{0}\to n  \pi^{0} $&$ \frac{\text{$\sin \theta$} \left(10 b_8-5 c_8-12 c_{27}-5d_8+10\mathbf{b^2_8}-12\mathbf{c^2_{27}}\right)}{40 \sqrt{3}}$&$35.9(5)$&$35.9(5)$&$0.692(17)$&$0.692(17)$\\\hline
		$\Sigma^{+}\to p  \pi^{0} $&$ \frac{\text{$\sin \theta$} \left(-5 c_8+5d_8\right)}{20 \sqrt{2}}$&$51.47(30)$&$50.90(24)$&$-0.982(14)$&$-0.99904(22)$ \\\hline
		$\Sigma^{+}\to n  \pi^{+} $&$ \frac{\text{$\sin \theta$} \left(5 d_8+5e_8\right)}{20 }$&$48.43(30)$&$48.50(29)$&$0.0489(26)$&$0.0486(26)$ \\\hline
		$\Sigma^{-}\to n  \pi^{-} $&$ \frac{\text{$\sin \theta$} \left(5 c_8+5e_8\right)}{20}$&$99.848(50)$&$99.849(50)$&$-0.0680(80)$&$-0.0706(76)$\\\hline
		$\Xi^{-}\to \Lambda^{0}  \pi^{-} $&$ \frac{\text{$\sin \theta$} \left(5a_8+5 b_8-10 c_8+4 c_{27}-10\mathbf{c^2_8}\right)}{20 \sqrt{6}}$&$99.887(35)$&$99.887(35)$&$-0.390(7)$&$-0.390(7)$ \\\hline
		$\Xi^{0}\to \Lambda^{0}  \pi^{0} $&$ \frac{\text{$\sin \theta$} \left(-5a_8-5 b_8+10 c_8+6c_{27}+10\mathbf{c^2_8}\right)}{40 \sqrt{3}}$&$99.524(12)$&$99.524(12)$ &$-0.349(9)$&$-0.3490(89)$\\
		\hline
	\end{tabular}
	\label{tab:su3_table}
\end{table*}

   For clearly seeing the contribution of $H_8$ and $H_{27}$ and the symmetry breaking term, we can factor out the corresponding wilson coefficient as\cite{Buchalla:1995vs}
\begin{eqnarray}
C_+(1{\rm GeV})=0.680,\;\;\;\;C_-(1{\rm GeV})=-2.164.
\end{eqnarray}
The factored form factor can be defined as 
\begin{eqnarray}
&&c^f_{27}=\frac{c_{27}}{C_+},\quad\quad\quad\;\;\mathbf{c^f_{27}}=\frac{\mathbf{c^2_{27}}}{C_+},\notag\\
&&A/B_8^f=\frac{A/B_8^f}{C_-},\quad q^f_8=\frac{q^f_8}{C_-},\;\;q=c,d,e.
\end{eqnarray}
Then one can see that the factored out SU(3) amplitude $c_{27}/C_+$ and $\mathbf{c^2_{27}}/C_+$ are smaller then the amplitude corresponding to $H_8$. 

However, since all amplitudes corresponding to $H_8$ contain both symmetric and symmetry-breaking contributions, it is difficult to isolate the effects of SU(3) symmetry breaking in these channels. Fortunately, the amplitude $\mathbf{c_{27}}$ only contains the symmetry breaking contribution. Therefore the ratio of form factor $R_f=\mathbf{f^c}_{27}/f^c_{27}$ and $R_g=\mathbf{g^c}_{27}/g^c_{27}$ can reflect the size of symmetry breaking as $R_f=-(75\pm5)\%$ and $R_g=-(742\pm 2141)\%$. We can conclude that the symmetry breaking effect in amplitude corresponding to $H_{27}$ at lest  $75\%$. However, we should also note that the form factor $g$ indicated these processes may contain a large symmetry breaking effect up to $\sim1000\%$. We strongly recommend that future experimental efforts aim to reduce the measurement uncertainties, so that the extent of SU(3) symmetry breaking can be determined with higher precision.

	\section{The other hyperon two-body decays induced by $s\to d$}
    Building upon the previous analysis of non-leptonic two-body decays of octet baryons, this section further extends the study to two-body decays of decuplet baryons and charmed baryons which induced by $s\to d$. Although these processes involve different types of particles, the Hamiltonian can also be treated with the same method. By decomposing the effective weak Hamiltonian into irreducible representations of SU(3), we can systematically construct the amplitude structures for various decay processes. 

    \subsection{The  decuplet  baryon two body decays}
    The light baryons decuplet $T_{10}$ can be written by flavor matrix as
    \begin{align}
    	&T_{10}^{111}=\Delta^{++},T_{10}^{112}=\frac{\Delta^{+}}{\sqrt{3}},T_{10}^{113}=\frac{\Sigma^{*+}}{\sqrt{3}},\nonumber\\
    	&T_{10}^{121}=\frac{\Delta^{+}}{\sqrt{3}},T_{10}^{122}=\frac{\Delta^{0}}{\sqrt{3}},T_{10}^{123}=\frac{\Sigma^{*0}}{\sqrt{6}},\nonumber\\ 
    	&T_{10}^{131}=\frac{\Sigma^{*+}}{\sqrt{3}},T_{10}^{132}=\frac{\Sigma^{*0}}{\sqrt{6}},T_{10}^{133}=\frac{\Xi^{*0}}{\sqrt{3}},\nonumber \\
    	&T_{10}^{211}=\frac{\Delta^{+}}{\sqrt{3}},T_{10}^{212}=\frac{\Delta^{0}}{\sqrt{3}},T_{10}^{213}=\frac{\Sigma^{*0}}{\sqrt{6}},\nonumber\\
    	&T_{10}^{221}=\frac{\Delta^{0}}{\sqrt{3}},T_{10}^{222}=\Delta^{-},T_{10}^{223}=\frac{\Sigma^{*-}}{\sqrt{3}},\nonumber\\
    	&T_{10}^{231}=\frac{\Sigma^{*0}}{\sqrt{6}},T_{10}^{232}=\frac{\Sigma^{*-}}{\sqrt{3}},T_{10}^{233}=\frac{\Xi^{*-}}{\sqrt{3}},\nonumber\\
    	&T_{10}^{311}=\frac{\Sigma^{*+}}{\sqrt{3}},T_{10}^{312}=\frac{\Sigma^{*0}}{\sqrt{6}},T_{10}^{313}=\frac{\Xi^{*0}}{\sqrt{3}},\nonumber\\
    	&T_{10}^{321}=\frac{\Sigma^{*0}}{\sqrt{6}},T_{10}^{322}=\frac{\Sigma^{*-}}{\sqrt{3}},T_{10}^{323}=\frac{\Xi^{*-}}{\sqrt{3}},\nonumber\\
    	&T_{10}^{331}=\frac{\Xi^{*0}}{\sqrt{3}},T_{10}^{332}=\frac{\Xi^{*-}}{\sqrt{3}},T_{10}^{333}=\Omega^-.
    \end{align}
    Using the decomposed Hamiltonian in Eq.\eqref{deha}, we can construct the decuplet  baryon two body decays as
    \begin{align}
    	\mathcal{M}_{10 \to 8 + 8}&=a_{27}(T_{10})^{ijk}(H_{27})_{\{ij\}}^{\{ln\}}(T_8)_{klm}P_n^m\nonumber\\
    	&+b_{27}(T_{10})^{ijk}(H_{27})_{\{ij\}}^{\{mn\}}(T_8)_{klm}P_n^l\nonumber\\
    	&+c_{27}(T_{10})^{ijm}(H_{27})_{\{ij\}}^{\{kn\}}(T_8)_{klm}P_n^l\nonumber\\
    	&+d_{27}(T_{10})^{ijn}(H_{27})_{\{ij\}}^{\{km\}}(T_8)_{klm}P_n^l\nonumber\\
    	&+e_{27}(T_{10})^{ikm}(H_{27})_{\{ij\}}^{\{ln\}}(T_8)_{klm}P_n^j\nonumber\\
    	&+f_{27}(T_{10})^{ikn}(H_{27})_{\{ij\}}^{\{lm\}}(T_8)_{klm}P_n^j\nonumber\\
    	&+a_{8}(T_{10})^{ijl}(H_{8})_{i}^{m}(T_8)_{\{jkl\}}P_m^k\nonumber\\
    	&+b_{8}(T_{10})^{ijm}(H_{8})_{i}^{k}(T_8)_{\{jkl\}}P_m^l\nonumber\\
    	&+c_{8}(T_{10})^{ijm}(H_{8})_{i}^{l}(T_8)_{\{jkl\}}P_m^k\nonumber\\
    	&+d_{8}(T_{10})^{ilm}(H_{8})_{i}^{j}(T_8)_{\{jkl\}}P_m^k.\label{M10}
    \end{align}
 Since the color symmetry of Hamiltonian can help us further constrain the number of independent SU(3) amplitude which is applied in octet baryon two body decays in previous section, after considering the color symmetry, the number amplitude corresponding to $H_{27}$ can only  be one as $e_{27}$ and the total amplitudes are
    \begin{align}
    	\mathcal{M}_{10 \to 8 + 8} &=e_{27}(T_{10})^{ikm}(H_{27})_{\{ij\}}^{\{ln\}}(T_8)_{klm}P_n^j\nonumber\\
    	&+a_{8}(T_{10})^{ijl}(H_{8})_{i}^{m}(T_8)_{\{jkl\}}P_m^k\nonumber\\
    	&+b_{8}(T_{10})^{ijm}(H_{8})_{i}^{k}(T_8)_{\{jkl\}}P_m^l\nonumber\\
    	&+c_{8}(T_{10})^{ijm}(H_{8})_{i}^{l}(T_8)_{\{jkl\}}P_m^k\nonumber\\
    	&+d_{8}(T_{10})^{ilm}(H_{8})_{i}^{j}(T_8)_{\{jkl\}}P_m^k.\label{eq35}
    \end{align}
    By expanding the expressions above, we obtain the decay amplitudes listed in Table \ref{table2}. Expressed by the IRA amplitude, we derive the following relations for amplitudes of decay channels as
    \begin{align}
    	\mathcal{M}(\Omega^{-}\to \Xi^0\pi^{-})&=\sqrt{3}\mathcal{M}(\Xi^{*0}\to \Sigma^+\pi^{-}).\label{MM}
    \end{align}
    
    The branching ratios of $T_{10} \rightarrow T_8P_8$ can be written as
    \begin{align}
    	\mathcal{B}(T_{10A} \to T_{8B}P_{8}) = \frac{\tau_{A} |p_{cm}|}{16 \pi m_{A}^{2}} \left| A(T_{10A} \to T_{8B}P_{8}) \right|^{2}.\label{BR}
    \end{align}
where $\tau_{A}$ is the lifetime of the initial decuplet baryon, $|p_{cm}|$ is the final-state momentum in the center-of-mass frame, and $A(T_{10A} \to T_{8B}P_{8})$ is the decay amplitude. From the measured branching ratio $ \mathcal{B}(\Omega^{-} \to \Xi^0 \pi^{-})=(24.3\pm0.7)\%$, the decay amplitude is extracted as $A(\Omega^{-} \to \Xi^0 \pi^{-}) = (9.66 \pm 0.15) \times 10^{-4}$. Using Eq.~\eqref{MM}, the decay amplitude for \( \Xi^{*0} \to \Sigma^+ \pi^{-} \) is derived from this result, and the corresponding branching ratio is then calculated via Eq.~\eqref{BR}:
\begin{align}
    \mathcal{B}(\Xi^{*0} \to \Sigma^+ \pi^{-}) = (8.04 \pm 0.51) \times 10^{-14}.
\end{align}
This branching ratio is found to be of the order \( \mathcal{O}(10^{-14}) \), consistent with theoretical expectations. The reason lies in the different dominant interactions: the \( \Omega^- \) decays primarily via weak interaction and has a relatively long lifetime of \( 8.21 \times 10^{-11}~\mathrm{s} \), whereas \( \Xi^{*0} \) decays mainly through strong interaction, leading to a much shorter lifetime on the order of \( \mathcal{O}(10^{-23})~\mathrm{s} \). Consequently, the resulting branching ratio for the weak decay of \( \Xi^{*0} \) is significantly suppressed, as reflected in the value obtained through Eq.~\eqref{BR}.

We consider the two decay channels $\Omega^{-} \to \Xi^0 \pi^{-}$ and $\Omega^{-} \to \Xi^- \pi^{0}$ to extract the parameters $a_8$ and $e_{27}$ by fitting to Eq.~\eqref{decaywidth}, using the corresponding branching ratios $\mathcal{B}(\Omega^{-} \to \Xi^0 \pi^{-}) = (24.3 \pm 0.7)\%$, $\mathcal{B}(\Omega^{-} \to \Xi^- \pi^{0}) = (8.55 \pm 0.33)\%$,
and the symmetric parameters $\alpha(\Omega^{-} \to \Xi^0 \pi^{-}) = 0.09 \pm 0.14$, $
\alpha(\Omega^{-} \to \Xi^- \pi^{0}) = 0.05 \pm 0.21$. Due to the large experimental uncertainties in the symmetric parameters $\alpha$ extracted from the decays \( \Omega^{-} \to \Xi^0 \pi^{-} \) and \( \Omega^{-} \to \Xi^- \pi^{0} \), these parameters are neglected in the present analysis. For analysis  the decay branching ratios, we can defined amplitude in Eq.\eqref{BR} of these two channels as 
\begin{eqnarray}
\left| A(\Omega^{-} \to \Xi^0 \pi^{-}) \right|&=&\frac{\sin \theta (5 a_8+4 e_{27})}{20},\notag\\
\left| A(\Omega^{-} \to \Xi^- \pi^{0}) \right|&=&\frac{\sin \theta (5 a_8-6 e_{27})}{20\sqrt{2}}.
\end{eqnarray}
Then the experimental data can be used to determined these amplitudes as 
\begin{align}
    a_8 = 0.3228 \pm 0.0037, \quad e_{27} = 0.027 \pm 0.0037.
\end{align}
 One can see that as we expected in previous discussion, the amplitude $e_{27}$ contributes less than $a_8$. 

With the fitted amplitude, we can give some predictions by extracting the parameter $D_8 = -b_8 + c_8 + 2d_8 = 1.8087 \pm 0.0094$  from the experimental data $ \mathcal{B}(\Omega^{-} \to \Lambda^0 K^{-}) = (67.7 \pm 0.7)\% $ with Eq.~\eqref{decaywidth}. This makes the IRA amplitudes for $\Xi^{*0} \to \Lambda^0 \pi^0 $ and $ \Xi^{*-} \to \Lambda^0 \pi^{-} $ become
\begin{eqnarray}
\left| A(\Xi^{*0} \to \Lambda^0 \pi^0) \right|&=&\frac{\sin \theta (15 a_8+5D_8-18 e_{27})}{120},\notag\\
\left| A(\Xi^{*-} \to \Lambda^0 \pi^{-}) \right|&=&\frac{\sin \theta (15 a_8+5D_8+12 e_{27})}{60\sqrt{2}}.
\end{eqnarray}
 However, due to an unknown phase angle $ \phi $ between $ D_8 $ and $ a_8, e_{27} $, the exact amplitudes for these two decay channels cannot be determined. Our results provide the range of branching ratios for different values of the phase angle $ \phi $.
 \begin{eqnarray}
4.59\times10^{-14}\leq\mathcal{B}(\Xi^{*0} \to \Lambda^0 \pi^0)\leq4.16\times10^{-13},\notag\\
4.87\times10^{-14}\leq\mathcal{B}(\Xi^{*-} \to \Lambda^0 \pi^{-})\leq8.65\times10^{-13}.\label{10br}
\end{eqnarray}
These predictions can be tested in future high-precision experiments.

    \subsection{The  charmed  baryon two body decay induced by $s\to d$}
  Induced by the $s\to d$ Hamiltonian the charmed baryon which contain the strange quark such as $\{\Xi_c^{\prime+},\;\Xi_c^{\prime0}\}$ can also decay into $\{\Xi^{+/0}_c,\;\Lambda_c\}$. Depending on whether the initial charmed baryon is the anti-triplet or the sextet baryon, the discussion can be further divided into two cases. The anti-triplet $T_{c\overline{3}}$ and sextet $T_{c6}$ under the SU(3) flavor symmetry can be written as
    \begin{eqnarray}\
    	T_{c\overline{3}}&=&
    	\begin{pmatrix}
    		0 & \Lambda_c^+ & \Xi_c^+ \\
    		-\Lambda_c^+ & 0 & \Xi_c^0 \\
    		-\Xi_c^+ & -\Xi_c^0 & 0 
    	\end{pmatrix},
    	T_{c6}=
    	\begin{pmatrix}
    		\Sigma^{++}_c & \frac{\Sigma_c^+}{\sqrt{2}} & \frac{\Xi_c^{\prime+}}{\sqrt{2}} \\
    		\frac{\Sigma_c^+}{\sqrt{2}} & \Sigma_c^0 & \frac{\Xi_c^{\prime0}}{\sqrt{2}} \\
    		\frac{\Xi_c^{\prime+}}{\sqrt{2}} &\frac{\Xi_c^{\prime0}}{\sqrt{2}} & \Omega^0_c 
    	\end{pmatrix}.\notag\\
    \end{eqnarray}
    
    Here the $SU(3)$ flavour anti-triplet charmed baryons can be also represented as $(T_{c\bar 3})_i=\epsilon_{ijk}(T_{c\bar 3})^{[jk]}=(\Xi_c^0,-\Xi_c^+,\Lambda^+_c)$.
     Following the same method, the irreducible representation amplitudes of decay $T_{c\bar 3}$ and $T_{c6}$ can be constructed.
\begin{table}
	\renewcommand{\arraystretch}{1.5}
	\caption{The irreducible representation amplitudes of the other hyperon two-body non-leptonic decay.}\label{table2}
	\begin{tabular}{|c|c|c|c|c|c|c|c|}
		\hline
		Channel & Amplitudes \\\hline
		$\Sigma^{*+}\to p\pi^{0} $&$ \frac{\sin \theta (-5 a_8+5b_8-5d_8+6 e_{27})}{20 \sqrt{6}}$ \\\hline
		$\Sigma^{*-}\to n\pi^{-} $ & $ \frac{\sin \theta (-5 a_8-5c_8-5d_8-4 e_{27})}{20 \sqrt{3}}$\\\hline
		$\Sigma^{*0}\to p\pi^{-} $ & $ \frac{\sin \theta (-5 a_8+5b_8-5d_8-4 e_{27})}{20 \sqrt{6}}$\\\hline
		$\Sigma^{*0}\to n\pi^{0} $ & $ \frac{\sin \theta (-5 a_8-5b_8-10c_8-5d_8+6 e_{27})}{40 \sqrt{3}}$\\\hline
		$\Omega^{-}\to \Xi^0\pi^{-} $ & $ \frac{\sin \theta (5 a_8+4 e_{27})}{20}$\\\hline
		$\Omega^{-}\to \Xi^-\pi^{0} $ & $ \frac{\sin \theta (5 a_8-6 e_{27})}{20\sqrt{2}}$\\\hline
		$\Xi^{*0}\to \Sigma^+\pi^{-} $ & $ \frac{\sin \theta (5 a_8+4 e_{27})}{20\sqrt{3}}$\\\hline
		$\Xi^{*0}\to \Sigma^0\pi^{0} $ & $ \frac{\sin \theta (5 a_8-5b_8-5c_8-6 e_{27})}{40\sqrt{3}}$\\\hline
		$\Xi^{*-}\to \Sigma^0\pi^{-} $ & $ \frac{\sin \theta (-5 a_8-5b_8-5c_8-4 e_{27})}{20\sqrt{6}}$\\\hline
		$\Xi^{*-}\to \Sigma^-\pi^{0} $ & $ \frac{\sin \theta (5 a_8+5b_8+5c_8-6 e_{27})}{20\sqrt{6}}$\\\hline
		$\Xi^{*0}\to \Lambda^0\pi^{0} $ & $ \frac{\sin \theta \left(15 a_8-5b_8+5c_8+10d_8-18 e_{27}\right)}{120}$\\\hline
		$\Xi^{*-}\to \Lambda^0\pi^{-} $ & $ \frac{\sin \theta (15 a_8-5b_8+5c_8+10d_8+12 e_{27})}{60\sqrt{2}}$\\\hline
		$\Xi^{*0}\to p K^{-} $&$ \frac{\sin\theta \left(b_8-d_8\right)}{4\sqrt{3}}$\\\hline
		$\Xi^{*-}\to nK^{-} $&$ \frac{\sin\theta \left(-c_8-d_8\right)}{4\sqrt{3}}$\\\hline
		$\Omega^{-}\to \Lambda^0 K^{-} $&$ \frac{\sin\theta\left(-b_8+c_8+2d_8\right)}{4\sqrt{6}}$\\
		\hline
		\hline
		$\Xi^{+}_c\to \Lambda^{+}_c\pi^{0} $ & $ \frac{\text{$\sin \theta$} \left(5 a_8-6 a_{27}\right)}{20 \sqrt{2}}$\\\hline
		$\Xi^{0}_c\to \Lambda^{+}_c\pi^{-} $ & $ \frac{\text{$\sin \theta$} \left(5 a_8+4 a_{27}\right)}{20}$\\
		\hline
		\hline
		$\Xi^{'+}_c\to \Lambda^{+}_c\pi^{0} $ & $ \frac{\text{$\sin \theta$} \left(-5 a_8-5 b_8+6 b_{27}\right)}{40}$\\\hline
		$\Xi^{'0}_c\to \Lambda^{+}_c\pi^{-} $ & $ \frac{\text{$\sin \theta$} \left(-5 a_8-5 b_8-4b_{27}\right)}{20\sqrt{2}}$\\\hline
		$\Omega^{0}_c\to \Xi^{0}_c\pi^{0} $ & $ \frac{\text{$\sin \theta$} \left(5 a_8-6 b_{27}\right)}{20\sqrt{2}}$\\\hline
		$\Omega^{0}_c\to \Xi^{+}_c\pi^{-} $ & $ \frac{\text{$\sin \theta$} \left(-5 a_8-4 b_{27}\right)}{20}$\\
		\hline
	\end{tabular}
	\label{tab:su3_table2}
\end{table}
    
    The amplitude for the two body decays of $T_{c\overline{3}}$ is given as
    \begin{align}
    	\mathcal{M}_{T_{c\overline{3}} \rightarrow T_{c\overline{3}}P_8} &= a_{27}(T_{c\overline{3}})_i(H_{27})^{\{il\}}_{\{jk\}}(T_{c\overline{3}})^jP_l^k\nonumber\\
    	&+a_{8}(T_{c\overline{3}})_i(H_{8})^k_j(T_{c\overline{3}})^jP_k^i.\label{33}
    \end{align}
    For excluding the color suppressed amplitude of these processes, the TDA amplitude are needed. The detailed discussion of TDA amplitude can be found in the appendix. A comparison with the topological diagrammatic approach (TDA) amplitudes reveals that the irreducible representation amplitudes (IRA) constructed in Eq.\eqref{33} are not subject to color suppression.

    The amplitude for the two body decays of $T_{c6}$ is given as
    \begin{align}
    	\mathcal{M}_{T_{c6} \rightarrow T_{c\overline{3}}P_8}& = a_{27}\left( T_{c6} \right)^{\{ij\}}\left( H_{27} \right)^{\{km\}}_{\{ij\}} \left( T_{\bar{c3}} \right)_{[kl]} P^l_m 
    	\nonumber\\
    	&+b_{27}\left( T_{c6} \right)^{\{ik\}} \left( H_{27} \right)^{\{lm\}}_{\{ij\}} \left( T_{\bar{c3}} \right)_{[kl]} P^j_m\nonumber\\
    	&+a_{8}\left( T_{c6} \right)^{\{ij\}} \left( H_{8} \right)^l_i \left( T_{\bar{c3}} \right)_{[jk]} P^k_l
    	\nonumber\\
    	&+b_{8}\left( T_{c6} \right)^{\{il\}} \left( H_{8} \right)^j_i \left( T_{\bar{c3}} \right)_{[jk]} P^k_l.
    \end{align}
    After excluding the color suppressed amplitude, we can get
    \begin{align}
    	\mathcal{M}_{T_{c6} \rightarrow T_{c\overline{3}}P_8}& =b_{27}\left( T_{c6} \right)^{\{ik\}} \left( H_{27} \right)^{\{lm\}}_{\{ij\}} \left( T_{\bar{c3}} \right)_{[kl]} P^j_m\nonumber\\
    	&+a_{8}\left( T_{c6} \right)^{\{ij\}} \left( H_{8} \right)^l_i \left( T_{\bar{c3}} \right)_{[jk]} P^k_l\nonumber\\
    	&+b_{8}\left( T_{c6} \right)^{\{il\}} \left( H_{8} \right)^j_i \left( T_{\bar{c3}} \right)_{[jk]} P^k_l.
    \end{align}
    Expanding the above equations, we will obtain the decay amplitudes given in Table \ref{table2}. 
    
	\section{Conclusion}

    In this work, we investigate the baryon two decay processed induced by $s\to d \bar u u$. Since the hyperon two body decay with $\Delta S=1$ are primarily driven by this Hamiltonian, which is the central focus of our study. Under the SU(3) flavor symmetry, the baryon two body decays can be easily expressed by several IRA amplitudes. For constructing the SU(3) irreducible amplitude, the Hamiltonian corresponding to $s\to d \bar u u$ must be decomposed as $3 \otimes 3 \otimes \overline{3} \otimes \overline{3} = 27 \oplus 10\oplus \overline{10} \oplus 8 \oplus 8 \oplus 8 \oplus 8 \oplus 1 \oplus 1$. Fortunately, under the symmetric and antisymmetric transformations, the expression of each irreducible representation can be derived in Eq.\eqref{decom}. After including the correlation of symmetry of quark field and anti-quark field, the decuplet Hamiltonian will automatically disappear and only octet and 27-plet can contribute.

    Using the decomposed Hamiltonian we can finally construct the IRA amplitude and by considering the color information in baryon state and Hamiltonian the number of independent amplitude can be reduced to six. As we expected, the flavor symmetry is not a good symmetry in  hyperon decays. Therefore we systemically introduced a symmetry breaking effect induced by the difference of quark mass. With the symmetry breaking effect, we can give a global analysis of the  hyperon two decay processes. Our analysis shows that the $Br(\Sigma^{+}\to p  \pi^{0} )$ and $\alpha(\Sigma^{+}\to p  \pi^{0} )$ have more than 1$\sigma$ deviation with our prediction and this puzzle can not be explained by SU(3) symmetry breaking effect, suggesting potential contributions from other mechanisms. Besides, We find that the symmetry breaking effect in amplitude corresponding to $H_{27}$ at lest  $75\%$ and our results indicated these processes may contain a large symmetry breaking effect about $1000\%$. We strongly recommend the experiment in the future can reduce the error of these data and the size of symmetry breaking can be accurately determined.
With the help of current experiment data, we can determine a part of amplitude and  the branching ratio of $\Xi^{*0} \to \Lambda^0 \pi^0$ and $\Xi^{*-} \to \Lambda^0 \pi^-$ can be estimated. However, due to the lack of experiment, the phase of amplitude can not be determined. Therefore we can only give the possible range of branching ratio in Eq.\eqref{10br} by varying its phase from $0$ to $\pi$. In the last part of this paper, we also analysis the charmed baryon two decays induced by $s\to d$. However, due to a lack of experimental data, we can only give the expression of amplitude.

	\section{Appendix}
	In section V, in order to simplify IRA further, we also investigate the TDA of the two-body decays of the charmed baryons in anti-triplet $T_{c\overline{3}}$ and the charmed baryons in anti-sextet $T_{c\overline{6}}$ caused by light quarks $s\rightarrow u\bar{u}d$. For the $s\rightarrow u\bar{u}d$ decays, the non-zero components of the effective Hamiltonian is
	\begin{align}
		&H_{13}^{12}=\sin \theta.
	\end{align}
	The amplitudes of processes $T_{c\overline{3}} \rightarrow T_{c\overline{3}}P_8$ and $T_{c\overline{6}} \rightarrow T_{c\overline{3}}P_8$ in topological diagrammatic approach(TDA) can be written as
	\begin{align}
		\mathcal{M}_{T_{c\overline{3}} \rightarrow T_{c\overline{3}}P_8} &=a_1(T_{c\overline{3}})^{[jk]}H_{ij}^{ml}(T_{c\overline{3}})_{[kl]}P_m^i\nonumber\\
		&+a_2(T_{c\overline{3}})^{[jk]}H_{ij}^{lm}(T_{c\overline{3}})_{[kl]}P_m^i\nonumber\\
		&+a_3(T_{c\overline{3}})^{[ij]}H_{ij}^{km}(T_{c\overline{3}})_{[kl]}P_m^l\nonumber\\
        &+a_4(T_{c\overline{3}})^{[jm]}H_{ij}^{kl}(T_{c\overline{3}})_{[kl]}P_m^i\nonumber\\
        &+a_5(T_{c\overline{3}})^{[jk]}H_{ij}^{im}(T_{c\overline{3}})_{[kl]}P_m^l\nonumber\\
        &+a_6(T_{c\overline{3}})^{[jm]}H_{ij}^{ik}(T_{c\overline{3}})_{[kl]}P_m^l.
	\end{align}
	\begin{align}
		\mathcal{M}_{T_{c\overline{6}} \rightarrow T_{c\overline{3}}P_8} &=b_1(T_{c\overline{6}})^{\{jk\}}H_{ij}^{ml}(T_{c\overline{3}})_{[kl]}P_m^i\nonumber\\
		&+b_2(T_{c\overline{6}})^{\{jk\}}H_{ij}^{lm}(T_{c\overline{3}})_{[kl]}P_m^i\nonumber\\
		&+b_3(T_{c\overline{6}})^{\{ij\}}H_{ij}^{km}(T_{c\overline{3}})_{[kl]}P_m^l\nonumber\\
        &+b_4(T_{c\overline{6}})^{\{jm\}}H_{ij}^{kl}(T_{c\overline{3}})_{[kl]}P_m^i\nonumber\\
		&+b_5(T_{c\overline{6}})^{\{jk\}}H_{ij}^{im}(T_{c\overline{3}})_{[kl]}P_m^l\nonumber\\
		&+b_6(T_{c\overline{6}})^{\{jm\}}H_{ij}^{ik}(T_{c\overline{3}})_{[kl]}P_m^l.
	\end{align}
	\begin{figure}[h!]
		\centering
		\includegraphics[width=1\linewidth]{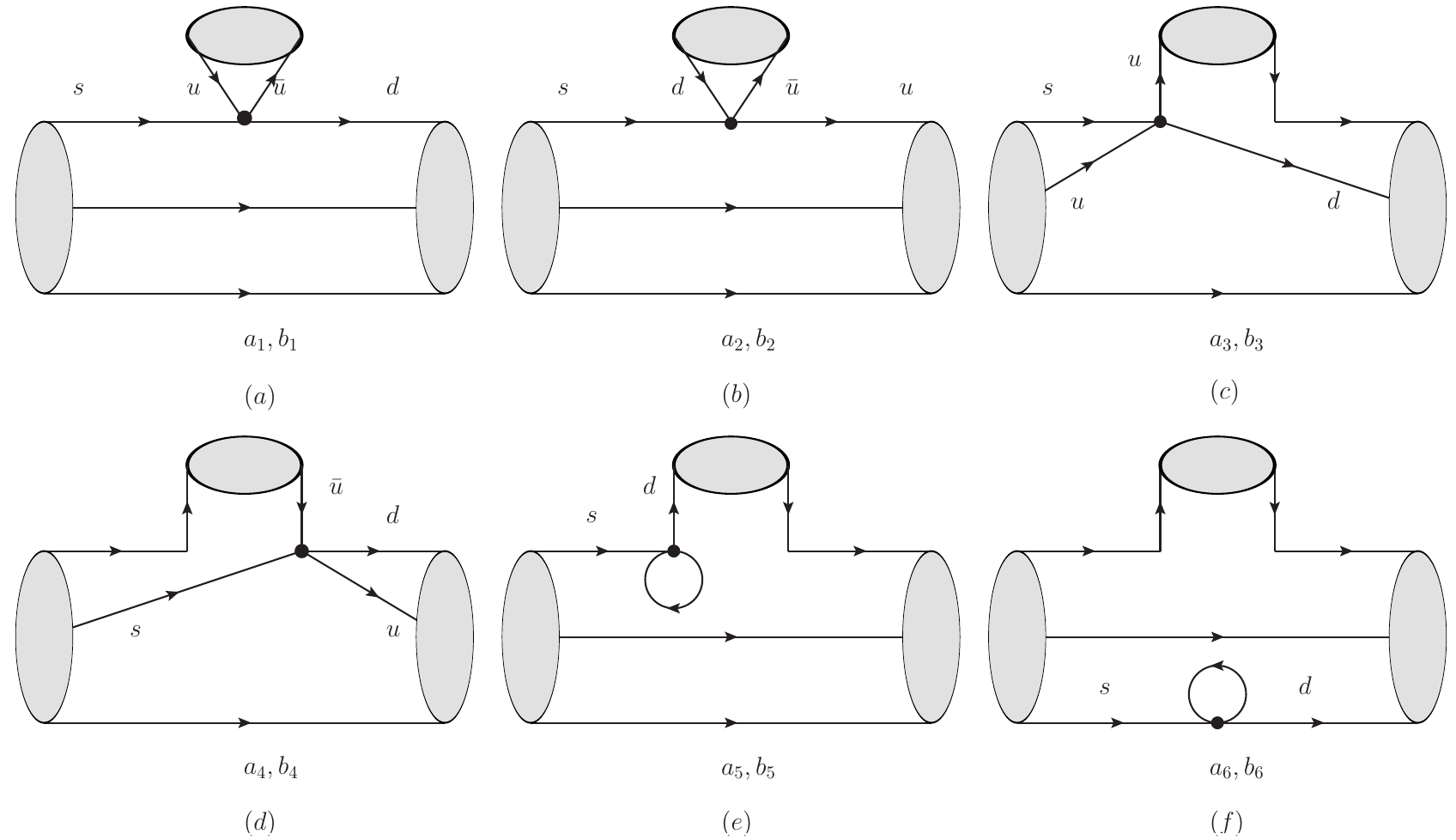}
		\caption{Topological diagrams for $T_{c\overline{3}} \rightarrow T_{c\overline{3}}P_8$ and $T_{c\overline{6}} \rightarrow T_{c\overline{3}}P_8$  nonleptonic decays.}
		\label{fig:antic}
	\end{figure}
	The relevant topological diagrams for $T_{c\overline{3}} \rightarrow T_{c\overline{3}}P_8$ and $T_{c\overline{6}} \rightarrow T_{c\overline{3}}P_8$ nonleptonic weak decays are displayed in Fig.2. The topological diagrams in Fig.2 can be divided into three categories: the tree diagrams
	in Fig.2 (a, b), the W-exchange diagrams in Fig.2 (c, d) and the penguin diagrams in Fig.2 (d, e).
	\section{Acknowledgements}
	We thank Dr.Xiao hui Hu in CUMT and Dr.Yuji Shi in ECUST for useful discussion.
The work of Ruilin Zhu is supported by NSFC under grant No. 12322503 and No. 12075124, and by Natural Science Foundation of Jiangsu under Grant No. BK20211267.
The work of Zhi-Peng Xing is supported by NSFC under grant No.12375088 , No. 12335003 and No. 12405113.
	The work of Ye Xing is supported by NSFC under grant No.12005294.

\end{document}